\documentclass[%
a4paper,
preprint,
showpacs, showkeys,
preprintnumbers,
amsmath,amssymb,
aps,
prb,
]{revtex4-1}

\usepackage{graphicx}
\usepackage{dcolumn}
\usepackage{bm}
\usepackage{ifpdf}

\newcolumntype{C}{>{$\displaystyle}c<{$}}
\newcolumntype{R}{>{$\displaystyle}r<{$}}

\usepackage{color}
\definecolor{darkblue}{rgb}{0,0,.6}

\ifpdf
\usepackage{epstopdf}
\usepackage[pdftex,unicode,pdfstartview={FitH},pdfborder={0 0 0}]{hyperref}
\usepackage{hypcap}
\else
\usepackage[hypertex]{hyperref}
\fi
\hypersetup{
	bookmarksnumbered = true,
	colorlinks = true, linkcolor = darkblue,
	citecolor = darkblue, filecolor = darkblue,
	menucolor = darkblue, urlcolor = darkblue
}

\let\vec=\mathbf
\DeclareMathOperator{\Ci}{Ci}
\DeclareMathOperator{\sinc}{sinc}

\begin{document}


\title{Giant Stokes shifts in AgInS$_2$ nanocrystals\\with trapped charge carriers}

\author{Anvar~S.~Baimuratov}
\email{anvar.baimuratov@lmu.de}
\affiliation{Fakult\"at f\"ur Physik, Munich Quantum Center, and Center for NanoScience (CeNS), Ludwig-Maximilians-Universit\"at M\"unchen, Geschwister-Scholl-Platz 1, D-80539 M\"unchen, Germany}

\author{Irina~V.~Martynenko}
\affiliation{Information Optical Technologies Centre, ITMO University, Saint Petersburg 197101, Russia}

\author{Alexander~V.~Baranov}
\affiliation{Information Optical Technologies Centre, ITMO University, Saint Petersburg 197101,
  Russia}

\author{Anatoly~V.~Fedorov}
\affiliation{Information Optical Technologies Centre, ITMO University, Saint Petersburg 197101, Russia}

\author{Ivan~D.~Rukhlenko}
\affiliation{Institute of Photonics and Optical Sciences (IPOS), School of Physics, The University of Sydney, Camperdown 2006, NSW, Australia}
\affiliation{Information Optical Technologies Centre, ITMO University, Saint Petersburg 197101, Russia}

\author{Stanislav~Yu.~Kruchinin}
\email{stanislav.kruchinin@univie.ac.at}
\affiliation{Center for Computational Materials Sciences, Faculty of Physics, University of Vienna, Sensengasse 8/12, 1090 Vienna, Austria}

\date{\today}

\begin{abstract}
Nanocrystals of AgInS$_2$ demonstrate giant Stokes shifts $\sim 1$~eV, the nature of which is still not clearly understood.
We propose a theoretical model of this phenomenon bringing together several different mechanisms previously considered only separately.
We take into account the contribution of electron-electron interaction with the hybrid density functional theory, as well as the renormalization of energy spectrum due to the electron-phonon coupling.
Furthermore, we consider the presence of at least one point defect responsible for hole trapping and the formation of a localized polaron state.
Our numerical simulations show that photoluminescence due to the recombination of a non-trapped electron and a trapped hole results in the giant Stokes shift in AgInS$_2$ nanocrystal, which is in close agreement with the recent experimental results.
\end{abstract}

\maketitle

\section{Introduction}\label{sec:intro}

In the last years, ternary I-III-VI (I = Cu, Ag; III = In, Sn, Ga, Al and VI = S, Se, Te) nanocrystals have received significant attention due to their compositional and structural versatility and unique optical properties.\cite{Bai2019, Girma2017, Kolny-Olesiak2013, Ulusoy2016, Xu2016}
These materials are promising candidates for the eco-friendly replacement of broadly studied II-VI and IV-VI binary nanocrystals containing inherently toxic elements such as Cd or Pb\cite{Pietryga2016, Xu2016}.
Particularly interesting examples of Cd-free ternary nanocrystals are CuInS$_2$ and AgInS$_2$ showing strong defect-state photoluminescence in the visible and near-infrared regions with comparatively high photoluminescence quantum yields exceeding 90\%.\cite{Bergren2018}
Their emission band is very broad, its characteristic FWHM (full width at half maximum) for nanocrystal ensembles is 300--800~meV and can be controlled by changing the composition, size, surface passivation, and ligand shell.\cite{Aldakov2013, Mao2011, Raevskaya2017, Song2016, Stroyuk2018, Stam2016}
The single-particle photoluminescence spectra recently measured for both CuInS$_2$ and AgInS$_2$ show significant broadening.
To some extent, this is an intrinsic property because the FWHM values of a single AgInS$_2$/ZnS nanocrystal are varying from 240 to 360~meV\cite{Stroyuk2019, Martynenko2019} while those of CuInS$_2$/ZnS nanocrystals are varying from 60~meV\cite{Zang2017} to 280~meV.~\cite{Whitham2016}
Another notable feature of the defect-assisted photoluminescence is its considerably long lifetime of a few hundreds of nanoseconds\cite{Aldakov2013, Chang2012, Deng2012, Raevskaya2017}, which makes AgInS$_2$ nanocrystals attractive for time-gated fluorescence, lifetime multiplexing and barcoding.\cite{Evstigneev2018}

The most striking characteristic feature observed experimentally in these materials is the giant Stokes shift.
Depending on the nanocrystal size and composition, this shift in CuInS$_2$ nanocrystals may vary from 200 to 500 meV\cite{Leach2016, Nagamine2018, Stam2016, Xia2018}, while AgInS$_2$ nanocrystals demonstrate even larger shifts between 300 and 1000~meV.\cite{Hamanaka2011, Jeong2017, Raevskaya2017, Stroyuk2019, Stroyuk2018}
Giant Stokes shifts open a possibility to achieve better efficiency in several technologies of light emission, including LEDs,\cite{Chen2018} solar cells,\cite{Chen2018, Suriyawong2016} and reabsorption-free luminescent solar concentrators.\cite{Hu2015, Klimov2016, Meinardi2015, Wu2018}
Indeed, recent studies by \textcite{Bergren2018} showed that a solar concentrator based on nearly reabsorption-free CuInS$_2$ nanocrystals that are spectrally tuned for optimal solar spectrum splitting have a far better performance than any analogues in solar concentrator technology.\cite{Bergren2018}
Furthermore, the giant Stokes shift makes ternary nanocrystals favorable for imaging of biological tissues.\cite{Martynenko2017}

Because of its technological relevance in optoelectronic and bioimaging applications, investigation of mechanisms responsible for the giant Stokes shift requires particular attention.
The broadband emission is a complex phenomenon which has recently became an object of intense debate in the literature.
Since the early days of the field, the donor-acceptor pair mechanism has been invoked to explain the radiative recombination in both AgInS$_2$ and CuInS$_2$ nanocrystals.\cite{Kolny-Olesiak2013, Stam2016, Hamanaka2011}
Recombination of a localized hole with a conduction band electron is the most likely emission mechanism for CuInS$_2$ nanocrystals, according to recent publications.\cite{Fuhr2017, Knowles2015, Nagamine2018}
\textcite{Zang2017} use the model of Cu-based defects, include quantum confinement effects and take into account the defect position for CuInS$_2$/ZnS.
For all the defect-based models of photoluminescence, the Stokes shift depends on binding energies of donor and/or acceptor trapping.

Another possible mechanism without defects, the exciton self-trapping, was proposed for ternary nanocrystals in Refs.~\onlinecite{Knowles2015, Raevskaya2017, Stroyuk2019, Stroyuk2018}.
It is based on an assumption of strong electron-phonon interaction and leads to a broadband emission even in the case of a single nanocrystal.
In this model, the magnitude of a Stokes shift increases due to a large number of emitted phonons.

\textcite{Shabaev2015} proposed a theory of photoluminescence from spherical chalcopyrite CuInS$_2$ nanocrystals explaining shifts of up to 300~meV between the first allowed and the first forbidden transitions.
In this approach, emission is generated by the formally forbidden transition and has a long photoluminescence lifetime.
Recently, \textcite{Nagamine2018} applied the two-photon absorption spectroscopy to confirm the existence of  two-photon transition below the single-photon band edge, which has never been observed before for any other semiconductor nanostructure.
This transition comes from the inversion of the $1S$ and $1P$ hole level order at the top of the valence band and results in a blue shift of the experimentally measured one-photon absorption edge by nearly 100--200~meV.
However, it is not large enough to explain the Stokes shift of 200--500~meV.

In this paper, we develop a unified model including several different mechanisms and apply it for simulation of giant Stokes shifts in AgInS$_2$ nanocrystals.
In particular, we take into account the purely electronic contributions using the density functional simulations and the $\vec k\cdot\vec p$-method~\cite{Kane_1956_JPCS_1_82, Luttinger1956} developed for tetragonal chalcopyrites in Ref.~\onlinecite{Shabaev2015}.
We consider the presence of at least one point defect, namely acceptor, and take into account the radial position of the defect.\cite{Zang2017, Fuhr2017, Knowles2015, Nagamine2018}
We assume that the acceptor localizes a hole in a relatively small area leading to the formation of a polaron state with LO (longitudinal optical) phonons.
The calculated polaron binding energy is large, similarly to the self-trapped and donor-acceptor pair models.\cite{Raevskaya2017, Stroyuk2019, Stroyuk2018, Hamanaka2011}
Finally, we calculate the size dependencies of Stokes shifts in AgInS$_2$ nanocrystals and explain their physical origins.
Our model predicts giant Stokes shifts of about 1~eV for nanocrystals with sizes of 1--2~nm, in agreement with the available experimental data.\cite{Hamanaka2011, Jeong2017, Raevskaya2017, Stroyuk2019, Stroyuk2018}

\section{Energy spectrum of A\lowercase{g}I\lowercase{n}S$_2$ nanocrystals}

\subsection{Electronic energies}

\begin{figure}[!htb]
  \centering\includegraphics[width=\textwidth]{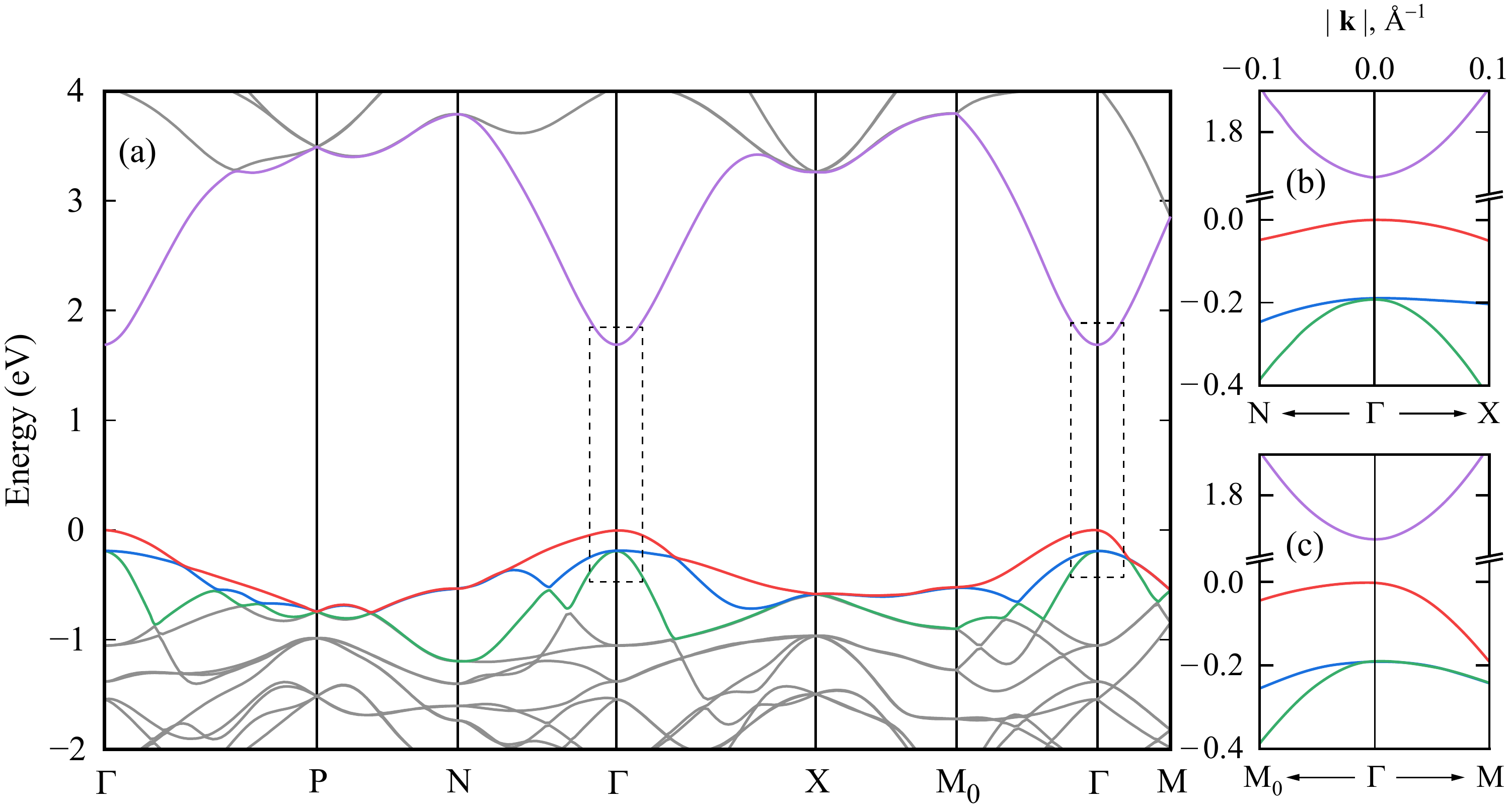}
  \caption{\label{f:01}%
    (Color online)
    (a) Energy bands of chalcopyrite AgInS$_2$ along the high-symmetry directions of the Brillouin zone.
    The bands included in the basis for calculation of the nanocrystal energy spectrum are highlighted by colors.
    [(b) and (c)] The close-up views of the areas outlined by dashed rectangles in (a).
  }
\end{figure}

\begin{figure}[!htb]
  \centering\includegraphics[width=0.5\textwidth]{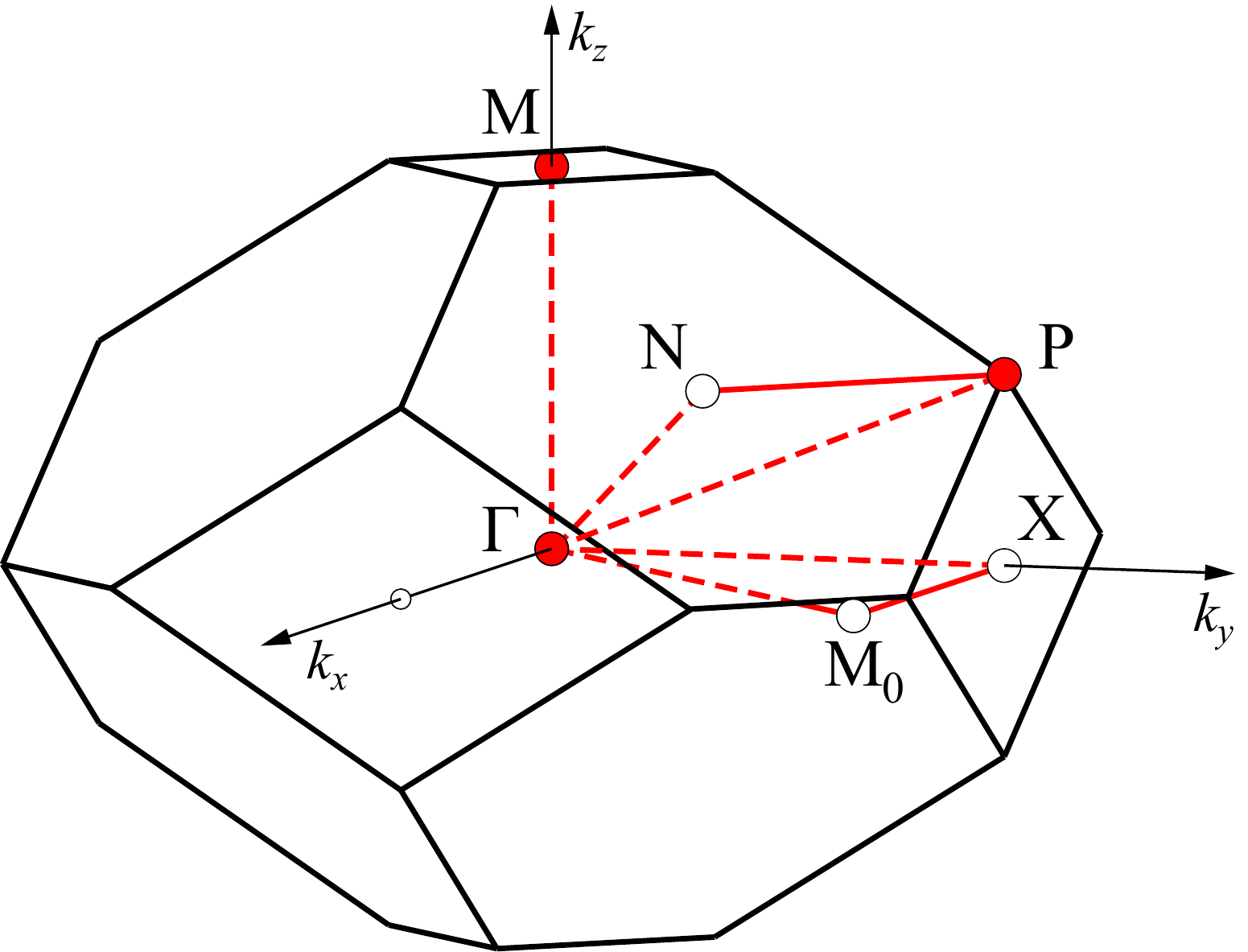}
  \caption{\label{f:02}%
    (Color online)
    Brillouin zone, high-symmetry directions and special points of chalcopyrite AgInS$_2$ (space group $I\bar 42d$, no. 122) in the direct lattice coordinates, which are defined according to Ref.~\onlinecite{Aroyo2011}.
    The orientation of reciprocal lattice vectors is obtained via clockwise rotation by $\pi/4$ around the $k_z$-axis.
  }
\end{figure}

\begin{figure}[!htb]
  \centering
  \includegraphics[scale=1]{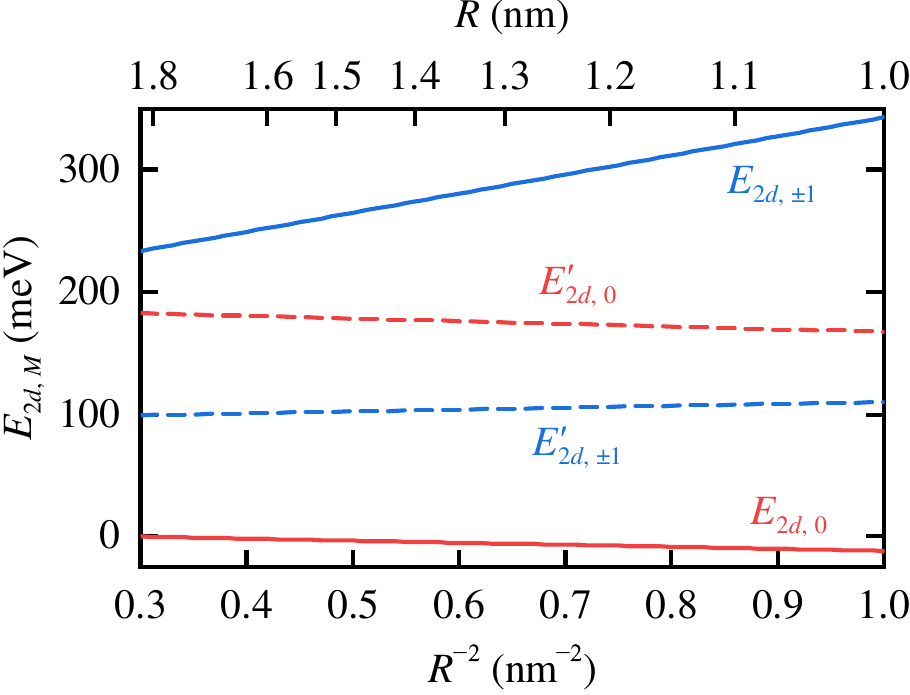}\hfill
  \caption{\label{fig03}%
  (Color online)
  Size dependencies of energy perturbations due to deviation of the lattice potential from the cubic symmetry for even (solid curves) and odd (dashed curves) hole states.
  }
\end{figure}

To describe electrons and holes confined in the AgInS$_2$ nanocrystal we use the density functional theory and the multiband $\vec{k} \cdot \vec{p}$-theory.~\cite{Luttinger1956}
The band structure and the Brillouin zone of AgInS$_2$ are shown in Figs.~\ref{f:01} and~\ref{f:02}, respectively.
As in the case of CuInS$_2$,~\cite{Shabaev2015} our simulations of AgInS$_2$ have shown that the spin-orbit splitting $\Delta_{s} = -25$~meV is much smaller than the energy difference between quantum-confined hole states ($\sim 100$~meV, see Fig.~\ref{fig03}), so we can use the spin unpolarized functionals.

The band structure was obtained from the \emph{ab initio} simulation with the \textsc{vasp} code~\cite{Kresse_1996_PRB_54_11169} and the B3LYP hybrid exchange functional~\cite{Becke_1993_JCP_98_5648}
\begin{equation*}
E_\mathrm{x} = E_\mathrm{x}^\mathrm{LSDA}
+ 0.2 E_\mathrm{x}^\mathrm{Fock}
+ 0.72 \Delta E_\mathrm{x}^\mathrm{GGA}.
\end{equation*}

The calculation was performed on a $\vec k$ grid of $6\times 6\times 6$ points with 120 bands.
The bands were interpolated with the \textsc{wannier90} program.\cite{Mostofi_2014_CPC_185_2309}
The calculated direct band gap at the $\Gamma$ point $E\mathrm{_g^{(B3LYP)}} = 1.69$~eV is close to the experimental one at the room temperature $E\mathrm{_g^{(exp)}} = 1.87$~eV \cite{You2002, Kameyama2015}.
In comparison to B3LYP, simulations with the HSE06 hybrid functional and G$_0$W$_0$ approximation yield similar effective masses and $\Delta_\mathrm{cs}$, but smaller values of the fundamental band gap: $E\mathrm{_g^{(HSE06)}} = 1.47$~eV and $E\mathrm{_g^{(G_0W_0)}} = 1.62$~eV.

The crystal-field splitting comes from the breakdown of the cubic symmetry in the chalcopyrite structure, and exists even in the absence of the spin-orbit coupling.\cite{Shay1974}
The crystal-field split band (cs) in AgInS$_2$ lies above the bands of heavy (hh) and light holes (lh).~\cite{Limpijumnong2002}
Thus the calculated splitting energy is negative, $\Delta_\mathrm{cs} = -187$~meV, and the order of valence bands is reverse to the one in CuInS$_2$.~\cite{Shabaev2015}

To calculate the Luttinger parameters,\cite{Luttinger1956} we fit the lowest conduction (c) and the highest valence bands (cs, hh, and lh) along the high-symmetry directions to points M, M$_0$, and X by parabolic functions of $\vec k$.
The close-up views for these directions in the vicinity of $\Gamma$ point are shown in Figs.~\ref{f:01}(b) and 1(c).
The effective masses found from the fitting procedure are summarized in Table~\ref{t:effmass}.
These masses are in close agreement with the previous DFT calculations.\cite{Huang2014, Liu2015}

\begin{table}[!htb]
  \caption{\label{t:effmass}%
    Effective masses of the lowest conduction band and three valence bands fitted from  a calculation with the B3LYP hybrid functional.
    All masses are given in the units of the free electron mass $m_0$.}
  \begin{tabular}{C|CCC}
    \hline\hline
    & \Gamma-\mathrm{M} & \Gamma-\mathrm{M_0} & \Gamma-\mathrm{X} \\ \hline
    m_\mathrm{c}  & 0.15    & 0.17    & 0.17   \\
    m_\mathrm{cs} & 0.19    & 0.83    & 0.83   \\
    m_\mathrm{hh} & 0.75    & 0.55    & 2.81   \\
    m_\mathrm{lh} & 0.75    & 0.18    & 0.14   \\
    \hline\hline
  \end{tabular}
\end{table}

In the the $\vec k \cdot \vec p$-method,~\cite{Kane_1956_JPCS_1_82} the electron energy is given by\cite{Efros2000, Shabaev2015}
\begin{equation}
  E_\mathrm{c} = \frac{\hbar^2 p^2}{2m_\mathrm{0}} \left(\alpha + \frac{E_\mathrm{P}}{E_\mathrm{g} + E_\mathrm{c}}\right),
\end{equation}
where $m_0$ is the mass of a free electron, $E_\mathrm{P}$ is the Kane energy, and $\alpha$ is the contribution of the remote bands to the electron effective mass.
The best fit of our first-principle spectrum is obtained for $E_\mathrm{P} = 8.8$~eV and $\alpha = 1.2$.
For $E_\mathrm{c} \ll E_\mathrm{g}$ the effective mass of electron is $m_\mathrm{c} = 0.16 m_0$.

In spherical nanocrystals of radius $R$ and an infinitely high confining potential, the energy and wave function of the electron on the ground $1S_e$ level can be written as follows:
\begin{equation}\label{e:en_electron}
  E_{1S_e} = \pi^2 E_R \left(\alpha + \frac{E_\mathrm{P}}{E_\mathrm{g} + E_{1S_e}}\right)
\end{equation}
and
\begin{equation}\label{e:wf_electron}
\psi_{1S_e}(\vec{r}) = \frac{1}{\sqrt{2\pi R}} \frac{\sin(\pi r/R)}{r},
\end{equation}
where $E_R = \hbar^2/(2 m_0 R^2)$.

The valence band structure can be constructed by the method of invariants introduced by Luttinger.\cite{Luttinger1956}
The parameters of this Hamiltonian were obtained from our \emph{ab initio} simulations.
Following \textcite{Shabaev2015}, we neglect the hole energy spectrum warping and find the confined valence band levels using the first-order perturbation theory with the spherically-symmetric Hamiltonian:
\begin{equation}\label{e:sphere}
H_\mathrm{sphere} = \frac{1}{2m_0} \big[(\gamma_1 + 4\gamma) p^2 - 6\gamma(\vec{pI})^2\big],
\end{equation}
where $\gamma = (2\gamma_2 + 3\gamma_3)/5$ (Ref.~\onlinecite{Lipari1970}), $\gamma_1$, $\gamma_2$, and $\gamma_3$ are the Luttinger parameters,\cite{Luttinger1956}
and $\vec{p}$ and $\vec{I}$ are the momentum and the spin-1 matrix operators, respectively.

The $D_{2d}$ point group symmetry of AgInS$_2$ allows for additional invariant terms $V_{2d}$, which can be added to the Hamiltonian in Eq.~\eqref{e:sphere}.
For the band structure obtained from our DFT calculation, we found that the valence band spectrum can be described by the Hamiltonian in Eq.~\eqref{e:sphere} with the following addition to the cubic Hamiltonian quadratic invariants:\cite{Limpijumnong2002}
\begin{multline}\label{e:v2d}
V_{2d} = -\Delta_\mathrm{cs} I_z^2 + \frac{1}{2m_0} \big[
(\gamma_{1\perp} + 4\gamma_{2\perp}) (p_x^2 + p_y^2)
- 6\gamma_{2\perp} (p_x^2 I_x^2 + p_y^2 I_y^2)\\
-12\gamma_{3\perp} \{p_x,p_y\} \{I_x,I_y\}
- 6\gamma_{4\perp} (p_x^2 I_y^2 + p_x^2 I_y^2)
\big],
\end{multline}
where $\{a,b\} = (ab + ba)/2$, $\gamma_1$, $\gamma_2$, $\gamma_3$ plus four additional Luttinger parameters $\gamma_{1\perp}$,
$\gamma_{2\perp}$, $\gamma_{3\perp}$, and $\gamma_{4\perp}$ are associated with the effective masses of the three valence bands in various crystallographic directions.

By applying the fitting procedure to our first-principle results, we obtain the data summarized in Table~\ref{t:Luttinger}.
Parameters $\gamma_1$, $\gamma_2$, and $\gamma_3$ are close to the parameters obtained for data fitting of CuInS$_2$ by \textcite{Shabaev2015}

\begin{table}[!tb]
  \caption{\label{t:Luttinger}%
    Luttinger parameters for AgInS$_2$ fitted from the simulation with the B3LYP hybrid density functional.
    We use notations introduced for chalcopyrite CuInS$_2$ by \textcite{Shabaev2015} and the fit parameter $\gamma_3$ from the non-parabolic spectrum along the $\Gamma - \mathrm{N}$ direction [see Fig.~\ref{f:01}(b)].
  }
  \begin{tabular}{C|CR}
    \hline\hline
    n& \gamma_n & \gamma_{n\perp} \\ \hline
    1 & 2.68 & 0.59 \\
    2 & 0.67 & 0.06 \\
    3 & 0.82  & 0.3 \\
    4 & -   & 0.1 \\
    \hline\hline
  \end{tabular}
\end{table}

In a spherical nanocrystal, one can separate variables for each state with total angular momentum $F$.
Here, we are interested only in the manifold of the lowest energy levels of holes with angular momentum $F=1$. Using techniques developed in Ref.~\onlinecite{Ekimov1986} we find that the wave functions of \textit{even} and \textit{odd} states with $F=1$ and angular momentum projection $M = \pm1,0$ can be written as\cite{Shabaev2015}
\begin{gather}\label{e:wf_even}
\psi_M(\vec{r}) = R_0(r) Y_{00}(\vartheta,\varphi) \chi_M
+ R_2(r)\sum_{\mu=-1}^1 C_{2,M-\mu;1,\mu}^{1,M}
Y_{2,M-\mu}(\vartheta,\varphi)\chi_\mu,
\end{gather}
and
\begin{equation}\label{e:wf_odd}
\psi'_M(\vec{r}) = R_1(r)\sum_{\mu=-1}^1
C_{1,M-\mu;1,\mu}^{1,M} Y_{1,M-\mu}(\vartheta,\varphi) \chi_\mu,
\end{equation}
where $Y_{lm}(\vartheta,\varphi)$ are the spherical harmonics,
$C_{lm;1,\mu}^{F,M}$ are the Clebsch-Gordan coefficients,
and $\chi_\mu$ are the spinors representing the eigenvectors of the $I_z$ operator.
Hereafter, all the primed symbols are related to the \textit{odd} hole states, while the non-primed symbols --- to the \textit{even} states.

Energies of these states without perturbation of potential $V_{2d}$ depend on Luttinger parameters and dimensionless coefficients $\phi$ and $\xi$.
Here, $\xi = 4.49$ is the first zero of the Bessel function $J_1(x)$ and $\phi = 4.57$ depends on the ratio $(\gamma_1 - 2\gamma)/(\gamma_1 + 4\gamma)$.
For \textit{even} and \textit{odd} states we have
\begin{subequations}
\begin{align}\label{e:en_even}
  \epsilon = E_R (\gamma_1 - 2\gamma) \phi^2
,\\\label{e:en_odd}
  \epsilon' = E_R (\gamma_1 - 2\gamma) \xi^2.
\end{align}
\end{subequations}

The fine structure of these states can be obtained as a perturbation associated with the deviation from cubic symmetry by the Hamiltonian in Eq.~\eqref{e:v2d}.
These energy perturbations, $E_{2d,M}$ and $E'_{2d,M}$, are determined as follows:
\begin{equation}
  E_{2d,M} = \int \mathrm{d}\vec{r}\,\psi^*_{M}(\vec{r}) V_{2d}  \psi_{M}(\vec{r}),
\end{equation}
where for $E'_{2d,M}$ one should replace $\psi$ by $\psi'$.

Finally, using the Luttinger parameters from Table~\ref{t:Luttinger}, we find the explicit values of perturbed energies $E_M = \epsilon + E_{2d,M}$ and $E'_M = \epsilon' + E'_{2d,M}$:
\begin{subequations}
  \begin{align}
  E_0       &= -0.03 \Delta_\mathrm{cs} + 0.98 \epsilon,\\
  E_{\pm1}  &= -0.99 \Delta_\mathrm{cs} + 1.17 \epsilon,\\
  E'_0      &=      -\Delta_\mathrm{cs} + 0.98 \epsilon',\\
  E'_{\pm1} &= -0.5  \Delta_\mathrm{cs} + 1.02 \epsilon'.
  \end{align}
\end{subequations}

Figure~\ref{fig03} shows energy corrections determining the fine structure of the hole states.
In contrast to the CuInS$_2$ nanocrystals, where the lowest energy state without the spin-orbit coupling is the optically inactive $E'_0$ state,\cite{Shabaev2015} the lowest energy state in AgInS$_2$ is the optically active $E_0$-state.
This is because the crystal splitting $\Delta_\mathrm{cs} = -187$~meV in AgInS$_2$ has an opposite sign and larger absolute value.
The order of other quantum-confined states is also different in these two chalcopyrite compounds and determined primarily by the crystal splitting.

\subsection{Polaron and Coulomb shifts}\label{sec:hamil}

The Hamiltonian of electronic and vibrational subsystems of a nanocrystal is given by~\cite{Kittel_1987, Rukhlenko_2011_OE_19_15459, Baimuratov_2014_OE_22_19707}
\begin{equation}\label{e:ham_sum}
  H = \sum_{p} E_p a_p^\dagger a_p^{\vphantom{\dagger}}
    + \sum_\vec{q} \hbar\Omega_{\vec q} b_{\vec q}^\dagger b_{\vec q}^{\vphantom{\dagger}}
    + H_\text{e-ph},
\end{equation}
where the first two terms correspond to the noninteracting electron-hole pairs and LO phonons, whereas the last one describes their interaction.
Creation (annihilation) of electron-hole pairs in the state $\psi_p = \psi_n \psi_m$ with the energy $E_p = E_n + E_\mathrm{g} + E_m + V_{p}$ is described by the operators $a_p^\dagger$ ($a_p^{\vphantom{\dagger}}$), where
$E_n + E_\mathrm{g}$ and $E_m$ are the electron and hole energies, respectively,
$E_\mathrm{g}$ is the band gap of the bulk semiconductor,
$V_{p}$ is the Coulomb interaction between the electron and hole.
LO phonons with the energy $\hbar\Omega_\vec{q}$ and wave vector $\vec q$ are described by the creation (annihilation) operators $b_\vec{q}^\dagger$ $(b_\vec{q}^{\vphantom{\dagger}})$.

The polar electron-phonon interaction in nanocrystals induces \emph{intraband} transitions between the states of electron-hole pairs.
We take into account only the diagonal part.
This allows representing the electron-phonon Hamiltonian as
\begin{equation}\label{e:Heph}
  H_\text{e-ph} =
  \sum_{p,\vec{q}} \sqrt{\frac{2 \pi e^2 \hbar \Omega_\vec{q}}{\varepsilon q^2 V}}
  \left(
       \rho^{(\vec{q})}_{p} a_p^\dagger a_p^{\vphantom{\dagger}} b_\vec{q}^{\vphantom{\dagger}} 
     + \mathrm{H.c.}
  \right),
\end{equation}
where $-e$ is the electron charge, $\varepsilon = (1/\varepsilon_\infty - 1/\varepsilon_0)^{-1}$, $\varepsilon_0$ and $\varepsilon_\infty$ are the reduced, low- and high-frequency dielectric permitivities of the bulk semiconductor, $V$ is the nanocrystal volume, and Fourier component of the charge carrier density is given by
\begin{equation}\label{e:rho_qp}
  \rho^{(\vec{q})}_{p} = \int\mathrm{d}\vec{r}
  \left(
    |\psi_{n} (\vec{r})|^2 - |\psi_{m}(\vec{r})|^2
  \right) e^{i\vec{qr}}.
\end{equation}
It should be noted that the spatial quantization of the optical vibrational eigenmodes can be easily done, but it gives qualitatively similar results.\cite{Fedorov1997}

We start from the elimination of the electron and hole coordinates by averaging over the envelopes of electrons and holes~\cite{Ipatova2001}
\begin{equation}\label{e:slow}
  H^{(p)} = E_p
  + \sum_\vec{q} \hbar\Omega_\vec{q} b_\vec{q}^\dagger b_\vec{q}^{\vphantom{\dagger}}
  + \sum_\vec{q} 
  \sqrt{\frac{2 \pi e^2 \hbar \Omega_\vec{q}}{\varepsilon q^2 V}}
  \left(
    \rho^{(\vec{q})}_{p}
  b_\vec{q}^{\vphantom{\dagger}}
  + \mathrm{H.c.} 
  \right).
\end{equation}

The unified consideration of interacting electrons and phonons leads to the formation of polarons.\cite{Pekar1983}
Their eigenstates can be found via diagonalization of the Hamiltonian~\eqref{e:slow}, $\widetilde H = U^\dagger H^{(p)} U$, with the Lee--Low--Pines transformation~\cite{Lee_1953_PR_90_297}
\begin{equation}\label{e:unitary}
  U = \exp \left[
  \sum_\vec{q}
  \sqrt{\frac{2 \pi e^2}{\varepsilon q^2 V \hbar \Omega_\vec{q}}}
  \left(
    \rho^{(\vec{q})}_{p}   b_\vec{q}^\dagger
  - \mathrm{H.c.} 
  \right)
  \right]
  .
\end{equation}

By replacing the summation over $\vec{q}$ by integration in~\eqref{e:unitary}, we find the energy of electron-hole pair $\mathcal{E}_p = E_p - \Delta_p$ renormalized by electron-phonon interaction, where the polaron binding energy 
is 
\begin{equation}\label{e:binding_general}
  \Delta_p = \frac{e^2}{2\varepsilon} \iint \,
  \frac{\mathrm{d}\vec{r} \mathrm{d}\vec{r'}}
  {|\vec{r} - \vec{r'}|} \big(|\psi_n (\vec{r})|^2 - |\psi_m (\vec{r})|^2\big)
  \big(|\psi_n (\vec{r'})|^2 - |\psi_m (\vec{r'})|^2\big).
\end{equation}

In the case of spherical nanocrystal of radius $R$, it is convenient to rewrite Eq.~\eqref{e:binding_general} in the dimensionless coordinates $\vec{x} \equiv \vec{r}/R$ and $\vec{y} \equiv \vec{r}'/R$ as
\begin{equation}\label{e:binding}
  \Delta_p = \frac{e^2}{2\varepsilon R} B_p,
\end{equation}
where
\begin{equation}\label{e:bp}
B_p = \iint\frac{\mathrm{d}\vec{x} \mathrm{d}\vec{y}}{|\vec{x} - \vec{y}|}
  \big(|\psi_n (\vec{x}R)|^2 - |\psi_m (\vec{x} R)|^2\big)
  \big(|\psi_n (\vec{y}R)|^2 - |\psi_m (\vec{y} R)|^2\big).
\end{equation}

The Coulomb interaction is considered as a perturbation to energy $E_{p}$
\begin{equation}
  V_p =-\frac{e^2}{\varepsilon_\infty R} C_p,
\end{equation}
where
\begin{equation}
  C_p = \iint\frac{\mathrm{d}\vec{x} \mathrm{d}\vec{y}}{|\vec{x} - \vec{y}|}
  |\psi_n (\vec{x} R)|^2| \psi_m (\vec{y} R)|^2.
\end{equation}

Using Eqs.~\eqref{e:wf_electron},~\eqref{e:wf_even}, and~\eqref{e:wf_odd}, we find the dimensionless parameters $B_p$ and $C_p$ as well as corrections $V_p$ and $\Delta_p$ to the energy spectrum of the electron-hole pair with $1S_e$ electron and holes of angular momentum $F=1$ in nanocrystals of radius $R = 1$~nm.
These results are summarized in Table~\ref{t:shifts}.

\begin{table}[!htb]
  \caption{\label{t:shifts}
    Dimensionless parameters $B_p$, $C_p$, polaron shift $\Delta_p$ and Coulomb shift $V_p$ of electron-hole pairs with $1S_e$ electron for AgInS$_2$ nanocrystals of radius $R = 1$~nm.
    Hereafter, we use the the experimental values of dielectric constants $\varepsilon_0 = 9.6$ and $\varepsilon_\infty = 6.7$ from Ref.~\onlinecite{Marquez1995}.
  }
  \begin{tabular}{CcC|CC|CC}
    \hline\hline
    m      & parity & M     & B_p & \Delta_p~\mathrm{(meV)} & C_p & V_p~\mathrm{(meV)} \\ \hline
    F = 1  & even   & 0     & 0.03  & 1 & 1.88  & -405\\
    F = 1  & even   & \pm 1 & 0.026 & 1 & 1.88  & -405\\
    F = 1  & odd    & 0     & 0.108 & 4 & 1.62  & -348\\
    F = 1  & odd    & \pm 1 & 0.081 & 3 & 1.62  & -348\\
    \hline\hline
  \end{tabular}
\end{table}

The polaron binding energy is relatively small, with the most significant value corresponding to the \textit{odd} hole state with angular momentum projection $M = 0$, whereas two \textit{even} hole states possess the lowest binding energies.
This is because \textit{even} hole states are determined by the $s$-type wave functions, while \textit{odd} hole states are formed by the wave functions of $p$-type.

\begin{figure}[!htb]
  \centering
  \includegraphics[scale=1]{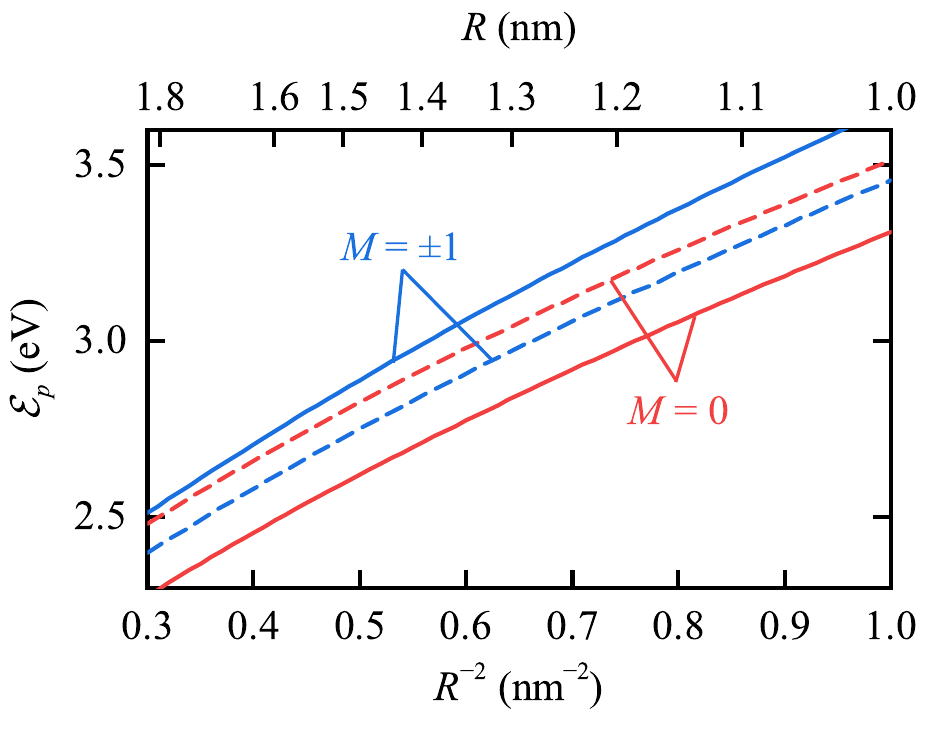}
  \caption{\label{fig04}%
    (Color online)
    Size dependencies of electron-hole pair energy in the lowest optically active state (solid) $\mathcal{E}_{1S_e,M}$ and optically passive state (dashed) $\mathcal{E}_{1S_e,M}'$ of AgInS$_2$ nanocrystal with the contribution of polaron binding energy $\Delta_{1S_e,M}$ and $\Delta_{1S_e,M}'$.
  }
\end{figure}

Figure~\ref{fig04} shows size dependencies of the electron-hole pair energies shifted by the Coulomb interaction and the polaron binding energy.
For all the considered electron-hole pairs with an electron in the $1S_e$ state the energy difference between their levels depends only on the energy of holes $E_m$ and shifts $V_{1S_e, M}$ and $\Delta_{1S_e, M}$.
Electron-hole pairs with \textit{even} hole states are optically active, while the pairs with \textit{odd} hole states are not.
The spin-orbit coupling is slightly activated optically by its admixture to the optically active \textit{even} hole level.\cite{Shabaev2015}
In any case, light absorption for optical transition between the state $1S_e$ and \textit{odd} hole states is rather weak.
Therefore, the lowest optically active level corresponds to the \textit{odd} hole state with $M=0$ and the lowest optically passive level is related to \textit{even} hole state with $M = \pm 1$.
We emphasize that the mechanism of Stokes shift proposed by \textcite{Shabaev2015} for CuInS$_2$ does not work for AgInS$_2$, due to the obtained reverse order of the lowest-energy levels (the lowest level is allowed in AgInS$_2$).

\subsection{Trap states}

Due to the structural complexity of ternary nanocrystals, many trap states are commonly observed.
The possible traps in these ternary compounds, namely donors and acceptors, are related to vacancies and interstitial atoms of Ag, In, and S,\cite{Hamanaka2011} or other charge defects such as substitution of In on the position of Ag and \emph{vice versa}.\cite{Chevallier2017}
If the nanocrystal has a shell of ZnS, there are additional mechanisms of defect formation \emph{via} Zn ions.\cite{Mao2013}

Without loss of generality, we consider an acceptor trap with charge $-Z$ ($Z$ is integer number).
We choose hole to be trapped charge carrier because its effective mass is much larger than that of an electron.
We assume that the trap states are described by binding energy $E_\mathrm{trap}(d)$ and the trial wave function
\begin{equation}\label{e:trap}
  \psi_\mathrm{trap}(\vec{r}-\vec{d}) =
  \begin{cases}
  \dfrac{\sin(\pi |\vec{r} - \vec{d}|/a)}{\sqrt{2\pi a |\vec{r} - \vec{d}|^2}} , & |\vec{r} - \vec{d}| \leqslant a;\\
  0, & |\vec{r} - \vec{d}| > a.
  \end{cases}
\end{equation}
where $\vec d$ is the trap position, and $a$ is the localization parameter of hole in the trap.

The total energy of an electron-hole pair in this state is given by
\begin{equation}
  E(d) = \mathcal{E}_{1S_e, M=0} - E_\mathrm{trap}(d).
\end{equation}

By assuming that the hole is strongly localized ($a \ll R$), and the trap state is bulk-like, \textit{i.e.} it does not strongly depend on the nanocrystal properties, we can write the binding energy of an electron-hole pair on the defect as follows:
\begin{equation}
  E_\mathrm{trap} (d) = E_\mathrm{bulk} - V(d) + \Delta_\mathrm{trap}(d),
\end{equation}
where the first term is the sum of the trapped hole kinetic energy and its Coulomb interaction with acceptor, the second term is the energy of the Coulomb interaction between the electron and the acceptor after trapping of the hole,
\begin{equation}
  V(d) = \frac{(Z-1) e^2}{\varepsilon_\infty R} f(d/R),
\end{equation}
$f(x) = 1 - \Ci(2 \pi) + \Ci (2 \pi x)- \ln x - \sinc(2 \pi x)$,
$\Ci(z)$ is the cosine integral,
and the last term is the polaron binding energy for the trap state
\begin{equation}
  \Delta_\mathrm{trap}(d) = \frac{e^2}{2\varepsilon R} B_\mathrm{trap} (d).
\end{equation}

\begin{figure}[!htb]
\centering
\includegraphics[height=5.7cm]{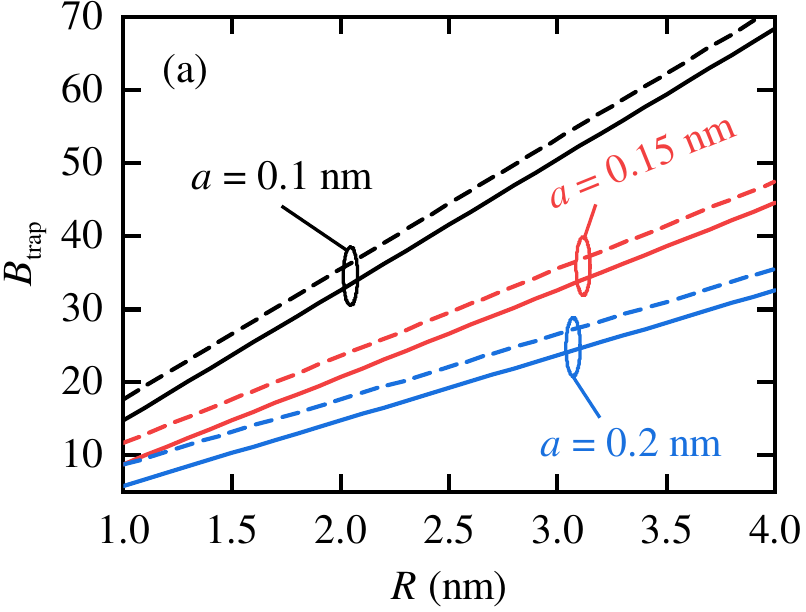}\hfill
\includegraphics[height=5.7cm]{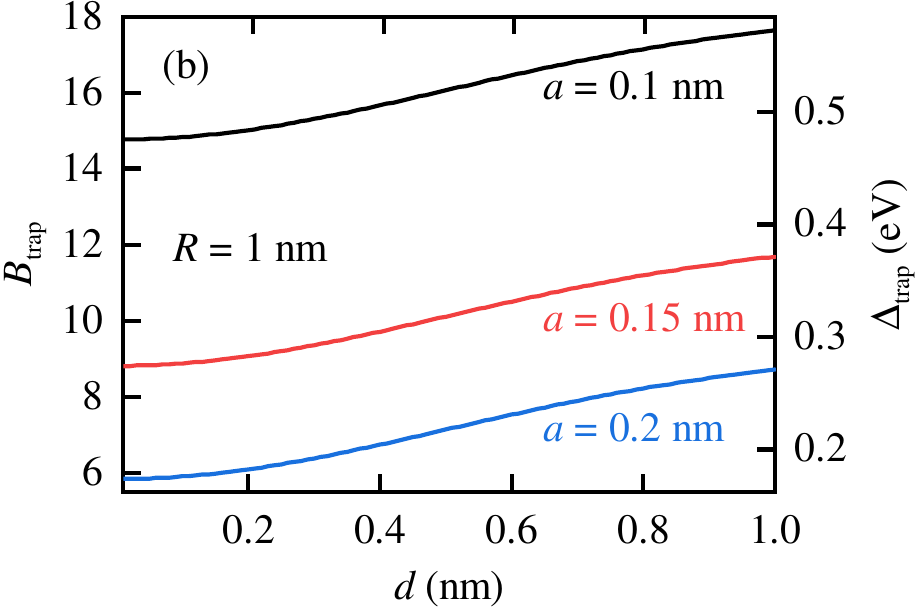}\hfill
\caption{\label{f:05}%
  (Color online)
  (a)~Size dependence of dimensionless strength of the polaron effect for states with trapped holes $B_\mathrm{trap} (d)$ for different defect positions $d = 0, R$ (solid and dashed lines, respectively).
  (b)~Dimensionless strength of the polaron effect $B_\mathrm{trap} (d)$ (left axis) and polaron binding energy $\Delta_\mathrm{trap} (d)$ (right axis) as functions of radial trap position in the nanocrystal of $R = 1$~nm for different values of the localization parameter $a$.
}
\end{figure}

From Eqs.~\eqref{e:bp},~\eqref{e:wf_electron}, and \eqref{e:trap} we find the strength of polaron effects for the states with trapped holes $B_\mathrm{trap} (d)$.
We plot this dimensionless strength in Fig.~\ref{f:05}, which shows that the obtained values of $B_\mathrm{trap} (d) \sim$~10--70 are by a few orders of magnitude larger than in the case of non-trapped states.
We calculate them for three different localization parameters, $a = 0.1, 0.15$, and $0.2$~nm and for two defect positions $d = 0, R$.
With the localization of a hole on the trap, the strength of polaron effect grows linearly due to the decrease of wave function overlapping of electron and hole.
Fig.~\ref{f:05}(b) shows that the maximal polaron strength relates to $d = R$.
For the estimation of polaron binding energy we use AgInS$_2$ nanocrystal of $R=1$~nm.
The obtained values of $\Delta_\mathrm{trap} (d)$ are shown on the right axis in Fig.~\ref{f:05}(b).

\section{Photoluminescence and Stokes shift}

\begin{figure}[!htb]
\centering
\includegraphics[width=0.49\textwidth]{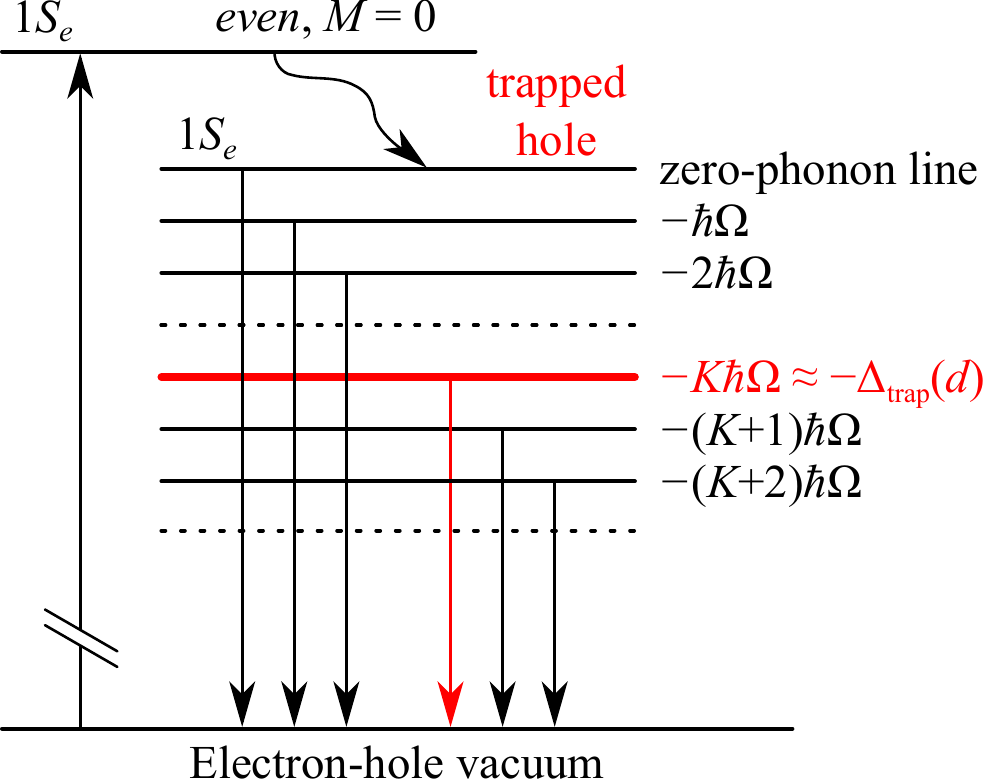}
\caption{\label{f:06}%
  (Color online)
  Schematic of absorption and photoluminescence from trapped hole states in AgInS$_2$ nanocrystals.
}
\end{figure}

Absorption and photoluminescence processes involving trapped states are schematically shown in Fig.~\ref{f:06}.
First, the absorbed light creates an electron-hole pair, which is relaxed to the lowest energy non-trapped state.
Second, the acceptor traps a hole, and finally, an electron recombines with the trapped hole and emits one photon and $N$ phonons.
In the approximation of dispersionless phonons, $\Omega_{\vec q} = \Omega$, the Stokes shift has a contribution from the double polaron binding energy because the phonon replica with the energy $K \hbar \Omega \approx \Delta_\mathrm{trap} (d)$ has the highest intensity in the photoluminescence spectrum.\cite{Ipatova2001} Here, $K$ is the Huang--Rhys parameter.\cite{Huang1950}
The probability of emission of photon of frequency $\omega$ and $N$ phonons can be calculated from the harmonic oscillator approximation\cite{landau1965quantum} 
\begin{equation}\label{e:rate}
  W(\omega) = \frac{4 \pi^2 e^2 \omega}{\varepsilon_\infty V} |r_{cv}|^2 |I (d)|^2
  \sum_N \frac{K^N e^{-K}}{N!} \delta \left[ \hbar\omega - E(d) + N \hbar \Omega \right],
\end{equation}
where $r_{cv}$ is the interband transition matrix element of the coordinate operator between the Bloch functions, which is given by $|r_{cv}|^2 = \hbar^2 E_\mathrm{P}/(m_0 E_\mathrm{g}^2)$, and $I (d) = \int \mathrm{d}\vec{r} \psi^*_{1S_e}(\vec{r}) \psi_\mathrm{trap}(\vec{r,d})$ is the overlap
integral of the electron and hole wave functions.

Assuming that the first absorption peak relates to the \textit{even} hole level with $M = 0$ and the electron in the $1S_e$-state and that the maximum of the photoluminescence spectrum relates to the $K$-phonon replica of the trapped state, we find the following expression for the Stokes shift of nanocrystal with a trapped hole
\begin{equation}\label{e:stokes}
  E_\mathrm{Stokes} (d) =  E_\mathrm{bulk} + E_0 + V_{1S_e, M=0} - V (d) - \left[ \Delta_{1S_e, M=0}-2 \Delta_\mathrm{trap} (d) \right].
\end{equation}

If we neglect small polaron shift $\Delta_{1S_e,M=0}$, we obtain
\begin{equation}\label{e:stokes_app}
  E_\mathrm{Stokes} (d) \approx  E_\mathrm{bulk} + E_0 + \frac{e^2}{R} \left[ \frac{(Z-1) f (d/R) - 1.88}{\varepsilon_\infty} + \frac{B (d)}{\varepsilon} \right].
\end{equation}
Notably, the Stokes shift depends on the kinetic term of the nontrapped hole $E_0$.

In recent experiments, the photoluminescence spectra from single nanocrystals of radius 1--2~nm demonstrate wide FWHM, $\gamma_S \approx 240$--360~meV.\cite{Stroyuk2019}
The phonon energy for this material is about 33~meV,\cite{Hamanaka2011}, which leads to quite a large number of $K$.

Assuming the finite linewidth of the replica and $K \gg 1$, we can rewrite Eq.~\eqref{e:rate} as
\begin{equation}\label{e:rate_mod}
  W(\omega) \approx \frac{4 \pi^2 e^2 \omega}{\varepsilon_\infty V} \frac{|r_{cv}|^2 |I (d)|^2}{\hbar \Omega \sqrt{2 \pi K}}
  \exp\left[ -\frac{1}{2 K} \left( \frac{E(d) - \hbar \omega -K \hbar \Omega}{\hbar \Omega}\right)^2 \right].
\end{equation}

The FWHM of this spectra is $\gamma_S \approx 2.355 \sqrt{K} \hbar \Omega$.
We estimate $K \sim 10$--21 and $\Delta_\mathrm{trap} (d) \sim 300$--700~meV by taking FWHM for a single nanocrystal as 240--360~meV from experimental data and find that the localization parameter $a$ is within the range of 0.08--0.14~nm.

\begin{figure}[!htb]
  \centering
  \includegraphics[width=0.49\textwidth]{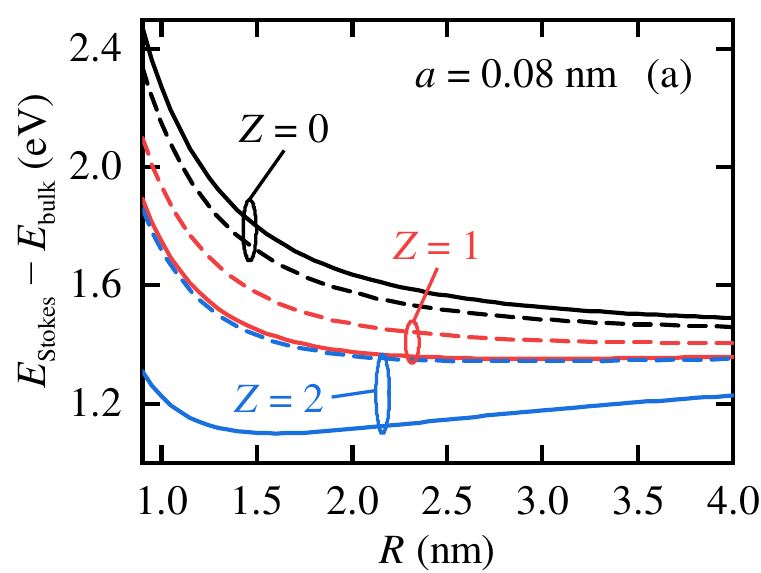}
  \includegraphics[width=0.49\textwidth]{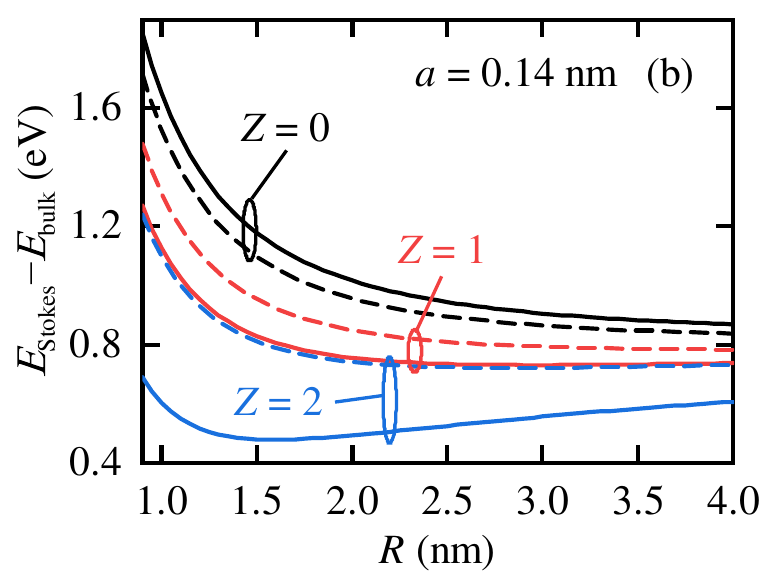}
  \includegraphics[width=0.49\textwidth]{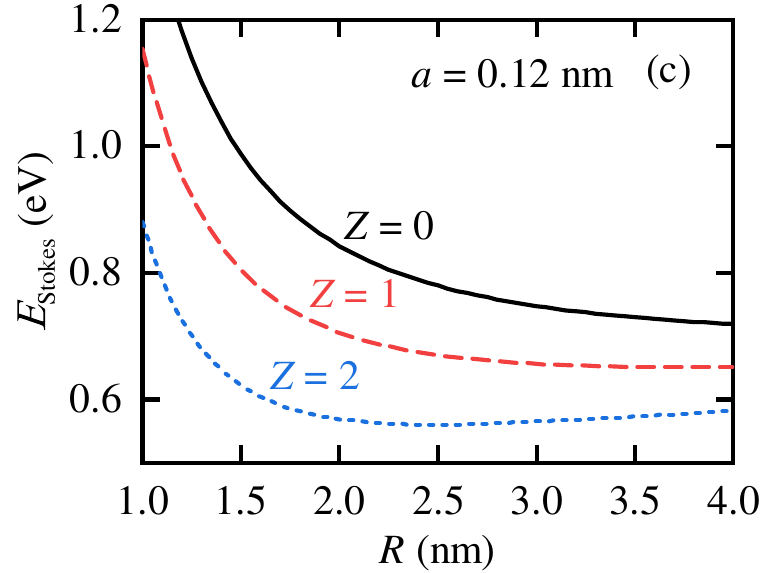}
  \includegraphics[width=0.49\textwidth]{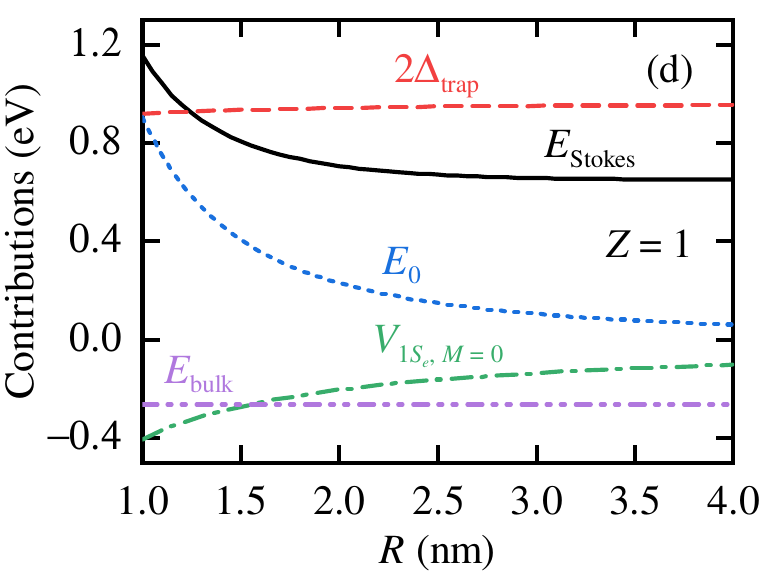}
  \caption{\label{f:07}%
    (Color online)
    Calculated Stokes shifts of AgInS$_2$ nanocrystals with trapped holes for localization parameters (a) $a = 0.08$~nm and (b) $a = 0.14$~nm.
    Holes are trapped by an acceptor at the center of nanocrystal ($d = 0$, solid curve) or on its surface ($d = R$, dashed curve).
    Panel~(c) shows the Stokes shifts for $a = 0.12$~nm, $d = 0.73 R$, and $E_\mathrm{bulk} = -260$~meV.
    (d)~Size dependence of individual contributions to the Stokes shift for a hole trapped by an acceptor with charge of $-1$.
    Other parameters are the same as in panel~(c).
  }
\end{figure}

Finally, to determine the Stokes shift we consider acceptors with charges $0,-1,-2$ (the noncharged acceptor with $Z = 0$ is related to the self-trapped model).
The results for Stokes shifts without a constant contribution $E_\mathrm{bulk}$ for defect positions $d = 0,R$ and the localization parameters $a = 0.08$~nm and $a = 0.14$~nm are shown in Figs.~\ref{f:07}(a) and \ref{f:07}(b).

To compare our numerically calculated Stokes shifts for nanocrystal ensembles with experimental data, we make a few additional assumptions.
First, we take the averaged value of the fitted data $a = 0.12$~nm as a localization parameter.
Second, for spherical nanocrystals the probability of finding a defect at position $d$ is proportional to $d^2$ whereas the overlap integral scales like $|I (d)|^2 \propto [d^2 \sin (\pi d/R)]^2$.
After taking the derivative of the product $d^2|I (d)|^2$, we solve the transcendental equation $\pi d_\mathrm{max} =-2 R\tan (\pi d_\mathrm{max}/R)$ and find that the largest contribution in ensemble comes from the nanocrystals with $d_\mathrm{max} = 0.73 R$.

To roughly estimate the value of $E_\mathrm{bulk}$ we assume the bulk limit $R \rightarrow \infty$ and find that $E_\mathrm{trap}^{(\infty)} = E_\mathrm{bulk} + 0.89 e^2/(\varepsilon a)$.
The binding energy of acceptor $E_\mathrm{trap}^{(\infty)}$ for bulk AgInS$_2$ was measured in several experiments.
For instance, the photoluminescence upon optical transitions with $E_\mathrm{trap}^{(\infty)} = 275$~meV was associated with a deep level of vacancies Ag, interstitial atoms of S,\cite{You2001} or other point defects $E_\mathrm{trap}^{(\infty)} = 180$--190~meV.\cite{Hamanaka2011}
We take the value $E_\mathrm{trap}^{(\infty)} = 220$~meV derived by \textcite{Hamanaka2011}, which is close to averaged value of the above mentioned experimental works.
This simple analysis gives $E_\mathrm{bulk} \approx -260$~meV.

The results of fitting for the acceptor with charges $0,-1,-2$ are shown in Fig.~\ref{f:07}(c).
For all three values of charge we find the giant Stokes shifts of 0.5--1~eV, which are in good agreement with experimental values.\cite{Hamanaka2011, Jeong2017, Raevskaya2017, Stroyuk2019, Stroyuk2018}
According to our simulations, the Stokes shift decreases with the QD size [see Fig.~\ref{f:07}(c)].
This dependence has been observed for II--VI and IV--VI binary nanoparticles, such as CdSe and CdS nanoparticles with excitonic emission. \cite{Efros1996, Yu2003}
\textcite{Stroyuk2018} demonstrated a similar dependency for aqueous glutation-capped size-selected AgInS$_2$/ZnS nanoparticles with diameters of 2--3 nm but explained it by assuming that the number of phonons involved in the optical process decreases with the size of nanoparticles.
On the other hand, the reported Stokes shifts of aqueous AgInS$_2$/ZnS QDs capped with mercaptoacetic acid are nearly constant in the 2.0--3.5~nm diameter range.\cite{Raevskaya2017}
This controversy between the experimental results could be explained by the insufficient accuracy of band gap measurements due to the absence of a sharp first peak in the absorption spectra. \cite{Martynenko2019}

Figure~\ref{f:07}(d) compares different contributions to the Stokes shift for a hole trapped by an acceptor with $Z = 1$.
There are two positive contributions: the doubled polaron binding energy $2\Delta_\mathrm{trap} (d_\mathrm{max})$ and the kinetic term of the non-trapped hole $E_0$.
The first one weakly grows with $R$ whereas the second one decreases as $\propto R^{-2}$.
The other two contributions are negative, the first one is related to $E_\mathrm{bulk}$, and the second one is the Coulomb interaction of the non-trapped state $V_{1S_e, M=0}$ decreasing with nanoparticle size as $\propto R^{-1}$.

Our model predicts that the size dependence of the Stokes shift is determined by two qualitatively different contributions.
The first one, $1.13 \hbar^2/(2 m_0 R^2)$, originates from the size-dependent part of a kinetic term $E_0$.
The second contribution, $e^2[(Z-1) f (d/R) - 1.88]/(\varepsilon_\infty R)$, is related to the Coulomb interaction.
The first term is always positive, whereas the sign of the second may change with charge parameter $Z$ and defect position $d$.
Remarkably, the polaron binding energy gives one of the main contributions to the Stokes shift and does not significantly dependent on the nanocrystal size.
Figure~\ref{f:07}(d) shows that only for small nanocrystals ($R \approx 1$~nm) the kinetic term $E_0$ closely  approaches the value of $2\Delta_\mathrm{trap}$.

\section{Conclusions}

We have developed a theoretical model of a giant Stokes shift in AgInS$_2$ nanocrystals with trapped charge carriers.
From the calculations with the B3LYP hybrid functional, we determined the Luttinger parameters and calculated the energies and wave functions of electrons and holes in spherical nanocrystals using the multiband $\vec k\cdot \vec p$-theory.~\cite{Kane_1956_JPCS_1_82, Luttinger1956, Shabaev2015}
For the first time, we calculated the size dependencies of the lowest electron-hole pairs with both Coulomb and polaron corrections.
We took into account the presence of one point defect acceptor trapping holes from the lowest non-trapped state.
The comparison of our results with experimental photoluminescence spectra for a single nanocrystal and the calculation of polaron binding energy for a trapped hole confirmed that the hole must be strongly localized with characteristic lengths of $a = 0.08$--0.14~nm.

Our simulations quantitatively reproduce the experimentally measured Stokes shift ($\sim 1$~eV) for a state with the hole localized on acceptors ($a = 0.12$~nm) with charges $0,-1,-2$ and radial position $d = 0.73 R$.
The main contribution is given by the polaron binding energy, $\Delta_\mathrm{trap} \sim 0.5$~eV.
The size dependence of the Stokes shift primarily originates from the kinetic term of initially non-trapped hole and Coulomb interaction in a confined quantum system.

\section{Acknowledgements}

A.\,S.\,B. has received funding from the European Union's Framework Programme for Research and Innovation Horizon 2020 (2014--2020) under the Marie Sk{\l}odowska--Curie Grant Agreement No. 754388 and from LMU Munich's Institutional Strategy LMUexcellent within the framework of the German Excellence Initiative (No. ZUK22).
I.\,V.\,M., A.\,V.\,B, A.\,V.\,F., and I.\,D.\,R. were partially funded by the Federal Target Program for Research and Development of the Ministry of Education and Science of the Russian Federation, grant no. 14.587.21.0047 (ID RFMEFI58718X0047).
S.\,Yu.\,K. acknowledges support from the Austrian Science Fund (FWF) within the Lise Meitner Project No. M2198-N30.
The numerical calculations were partially performed at the Vienna Scientific Cluster (VSC-3). 

\bibliography{bibliography}

\end{document}